\documentclass[twocolumn,aps,prc,showpacs,floatfix]{revtex4}
\usepackage{graphicx}
\usepackage{dcolumn}
\usepackage{bm}

\begin{document}

\title{Normal or abnormal isospin-fractionation as a qualitative
probe of nuclear symmetry energy at supradensities}

\author{Wenmei Guo$^{1,2,3,4}$}
\author{Gaochan Yong$^{1,6,7}$}\email{yonggaochan@impcas.ac.cn}
\author{Yongjia Wang$^{4,5}$}
\author{Qingfeng Li$^{4}$}
\author{Hongfei Zhang$^{5,6,7}$}
\author{Wei Zuo$^{1,6,7}$}
\affiliation{%
$^1${Institute of Modern Physics, Chinese Academy of Sciences, Lanzhou 730000, China}\\
$^2${School of Physical Science and Technology, Lanzhou
University, Lanzhou 730000, China}\\
$^3${University of Chinese Academy of Sciences, Beijing 100049, China}\\
$^4${School of Science, Huzhou University, Huzhou 313000, China}\\
$^5${School of Nuclear Science and Technology, Lanzhou University, Lanzhou 730000, China}\\
$^6${State Key Laboratory of Theoretical Physics, Institute of
Theoretical Physics, Chinese Academy of Sciences, Beijing,
100190}\\
$^7${Kavli Institute for Theoretical Physics, Chinese Academy of
Sciences, Beijing 100190, China}\\
}%

\date{\today}

\begin{abstract}
Within two different frameworks of isospin-dependent transport
model, effect of nuclear symmetry energy at supradensities on the
isospin-fractionation (IsoF) was investigated. With
positive/negative symmetry potential at supradensities (i.e.,
values of symmetry energy increase/decrease with density above
saturation density), for energetic nucleons, the value of neutron
to proton ratio of free nucleons is larger/smaller than that of
bound nucleon fragments. Compared with extensively studied
quantitative observables of nuclear symmetry energy, the normal or
abnormal isospin-fractionation of energetic nucleons can be a
qualitative probe of nuclear symmetry energy at supradensities.
\end{abstract}


\maketitle


The symmetry energy, which governs phenomena from the structure of
exotic nuclei to astrophysical processes, has many ramifications
in both nuclear physics and astrophysics
\cite{JM.Lattimer:2004,Steiner:2005,VBaran:2005,LiBA:08} and also
the study of Gravitational waves \cite{Fattoyev:2014}.
Unfortunately, nowadays predictions on nuclear symmetry energy
especially at supra-saturation densities are significantly
different for different many-body theory approaches
\cite{chenf07}.
\begin{figure}[t!]
\begin{center}
\includegraphics[width=0.5\textwidth]{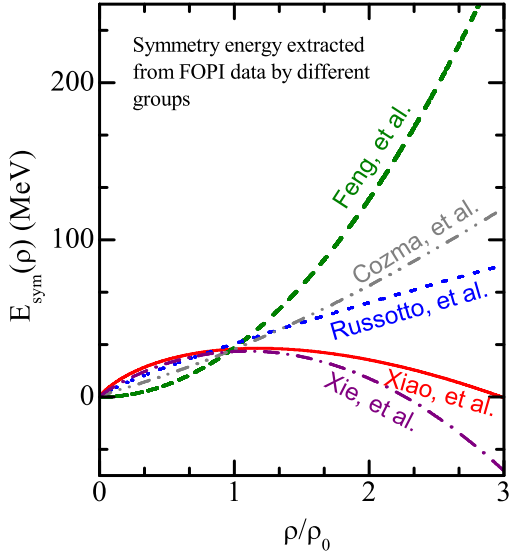}
\end{center}
\caption{Density dependent nuclear symmetry energy extracted from
FOPI and FOPI-LAND data by different groups. Xiao \emph{et al.}
made IBUU04 calculations and compared with FOPI pion data
\cite{XiaoZG:2009}. Feng \emph{et al.} made LQMD calculations and
also compared with FOPI pion data \cite{FengZQ:2010}. Russotto
\emph{et al.} made UrQMD calculations and compared with FOPI-LAND
nucleon elliptic flow data \cite{Russotto2011}. Xie \emph{et al.}
made Boltzmann-Langevin model calculations and compared with FOPI
pion data \cite{xie2013}. Cozma \emph{et al.} made T\"{u}bingen
quantum molecular dynamics (QMD) model calculations and compared
with FOPI-LAND elliptic flow data \cite{cozma2013}. } \label{esym}
\end{figure}

Although nuclear symmetry energy and its slope at normal density
of nuclear matter from recent 28 analyses of terrestrial nuclear
laboratory experiments and astrophysical observations have been
roughly pinned down \cite{lihan13}, recent interpretations of FOPI
\cite{Reisdorf2010} and FOPI-LAND \cite{Russotto2011,cozma2013}
data by different transport models give divergent
density-dependent symmetry energy at supradensities
\cite{XiaoZG:2009,FengZQ:2010,Russotto2011,xie2013,cozma2013}.
Divergence is shown in figure~\ref{esym}. One sees that the
constraints from elliptic flow obtained by using different models
(UrQMD and T\"{u}bingen QMD, as well as the newly updated version
of UrQMD model, see refs.~\cite{Russotto2011,cozma2013,wangy14})
are quite consistent among them, whereas the three constraints
from $\pi^{-}/\pi^{+}$ diverges
\cite{XiaoZG:2009,FengZQ:2010,xie2013}. In fact the use of
$\pi^{-}/\pi^{+}$ ratio to constraint nuclear symmetry energy
raises several doubts: Pion has large freeze-out time, delta and
pion scattering and re-absorption may destroy the high density
signals \cite{ditoro06}. Treatment of delta dynamics and
isospin-dependent pion in-medium effects in transport models is
not so straightforward \cite{xu2010,xu2013,ko2014}. Recent work of
MSU groups demonstrates that the ratio of pions spectra is more
sensitive than ratios of integrated yields \cite{Lynch2014}.
However, they did not distinguish pions messenger of high density
from the rest \cite{liu2014}. Moreover the super-soft behavior of
Xie \emph{et al} \cite{xie2013} and Xiao \emph{et al}
\cite{XiaoZG:2009} is not fairly compatible with neutron star
stability and structure except introducing non-Newtonian gravity
\cite{wen2009}.

%
Putting $\pi^{-}/\pi^{+}$ ratio aside, based on two different
transport models we propose isospin-fractionation of energetic
nucleons and fragments as a qualitative probe of nuclear symmetry
energy in heavy-ion collision at intermediate energies. We find
that with very soft symmetry energy, abnormal
isospin-fractionation in heavy-ion collision at intermediate
energies occurs, whereas with the stiff symmetry energy normal
isospin-fractionation is obtained. If having kinetic energy
distribution of free and bound nucleons data from, e.g, central
Au+Au reaction at 400 MeV/nucleon (as done by FOPI-LAND
Collaboration \cite{Russotto2011,cozma2013}), then one can
qualitatively determine high-density behavior of nuclear symmetry
energy.


\begin{figure*}[htb]
\centering\emph{}
\includegraphics[width=0.99\textwidth]{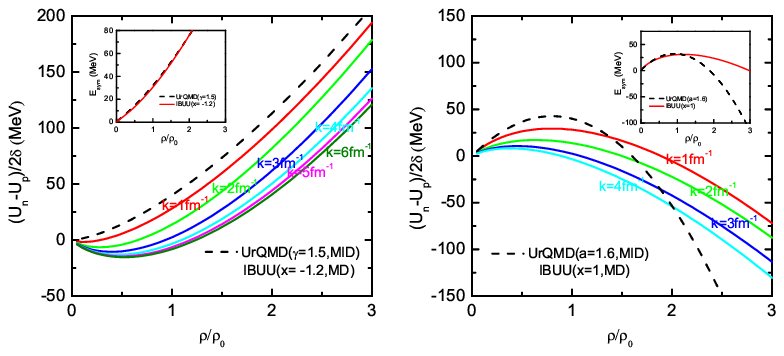}
\caption{Density dependent nuclear symmetry potential used in
IBUU04 and isospin-dependent UrQMD models. The lines labelled by
different momenta (MD) are symmetry potentials used in the IBUU04
model. Symmetry potentials used in present UrQMD model is momentum
independent (MID). Left window shows positive potential while
Right window shows negative potential, which respectively
correspond stiff and soft nuclear symmetry energies at
supra-saturation densities.} \label{potential}
\end{figure*}
In our used IBUU04 model, besides isospin-dependent initialization
and pauli-blocking, an isospin- and momentum-dependent mean-field
potential is adopted, i.e.,
\begin{eqnarray}
U(\rho,\delta,\textbf{p},\tau)&=&A_u(x)\frac{\rho_{\tau'}}{\rho_0}+A_l(x)\frac{\rho_{\tau}}{\rho_0}\nonumber\\
& &+B(\frac{\rho}{\rho_0})^{\sigma}(1-x\delta^2)-8x\tau\frac{B}{\sigma+1}\frac{\rho^{\sigma-1}}{\rho_0^\sigma}\delta\rho_{\tau'}\nonumber\\
& &+\frac{2C_{\tau,\tau}}{\rho_0}\int
d^3\,\textbf{p}'\frac{f_\tau(\textbf{r},\textbf{p}')}{1+(\textbf{p}-\textbf{p}')^2/\Lambda^2}\nonumber\\
& &+\frac{2C_{\tau,\tau'}}{\rho_0}\int
d^3\,\textbf{p}'\frac{f_{\tau'}(\textbf{r},\textbf{p}')}{1+(\textbf{p}-\textbf{p}')^2/\Lambda^2},
\label{buupotential}
\end{eqnarray}
where $\delta=(\rho_n-\rho_p)/(\rho_n+\rho_p)$ is the isospin
asymmetry, and $\rho_n$, $\rho_p$ are the neutron ($\tau=1/2$) and
the proton ($\tau=-1/2$) densities, respectively, and $\tau \neq
\tau^{'}$, $\sigma = 4/3$, $f_{\tau}(\textbf{r},\textbf{p})$ is
the phase-space distribution function at coordinate \textbf{r} and
momentum \textbf{p}. The variable $x$ is used to mimic different
forms of the symmetry energy/potential. More details can be found
in the recent paper \cite{gwm2013}. In fact within transport
model, the kinetic part of the symmetry energy is simulated by
using different Fermi momenta for neutrons and protons according
to the local Thomas-Fermi approximation while the potential part
is taken into account by using the symmetry potential. Uncertainty
of symmetry energy at supradensities thus comes from the
model-dependent symmetry potential. Shown in Fig.~\ref{potential}
is nuclear symmetry potential/symmetry energy used in the
transport models (together with symmetry potential/symmetry energy
used in the UrQMD model). In the IBUU04 model, we employed
parameters $x=-1.2$ and $x=1$ to represent the positive and the
negative symmetry potentials at supradensities, respectively.

In our used isospin-dependent UrQMD transport model, the potential
is expressed as \cite{LIQF:11}
\begin{eqnarray}
U&=& U^{(2)}_{Sky}+U^{(3)}_{Sky}+U_{Yuk}+U_{Cou}\nonumber\\
& &+U_{Pau}+U_{md}+U_{sym}.
\end{eqnarray}
The symmetry potential energy density used in the UrQMD transport
model is
\begin{eqnarray}
W_{sym}=(S_0-\frac{\epsilon_{F}}{3})\rho\cdot F(u)\cdot \delta^2,
\end{eqnarray}
here $S_0$= 32 MeV is the symmetry energy at the normal nuclear
density $\rho_0$. $\epsilon_{F}$ is the Fermi kinetic energy at
normal nuclear density. $u=\rho/\rho_0$ is the reduced nuclear
density, and $\delta=(\rho_{n}-\rho_{p})/(\rho_{n}+\rho_{p})$ is
isospin asymmetry. The symmetry potential is given by $
U^{n(p)}_{sym}=\partial{W_{sym}}/\partial{\rho_{n(p)}} $. For the
density dependent part $F(u)$, we use two forms \cite{liq2005}:
\begin{eqnarray}
F(u)=\left\{\begin{array}{l@{\quad,\quad}l}F_1=u^{\gamma} & \gamma>0
\\F_2=u\frac{a-u}{a-1} & a>1 \end{array} \right..
\end{eqnarray}
Here parameters $\gamma$ and $a$ are used to describe the density
dependence of symmetry potential. In figure~\ref{potential}, we
choose $\gamma= 1.5$ and $a= 1.6$ to describe positive and
negative symmetry potentials at supradensities, respectively. We
thus adopted roughly consistent forms for the positive or negative
symmetry potentials in the used two transport models. As for the
two-body scattering cross sections in medium, we used the same
in-medium corrected forms as those in Ref.~\cite{gwm2013}.


\begin{figure*}[htb]
\centering\emph{}
\includegraphics[width=0.99\textwidth]{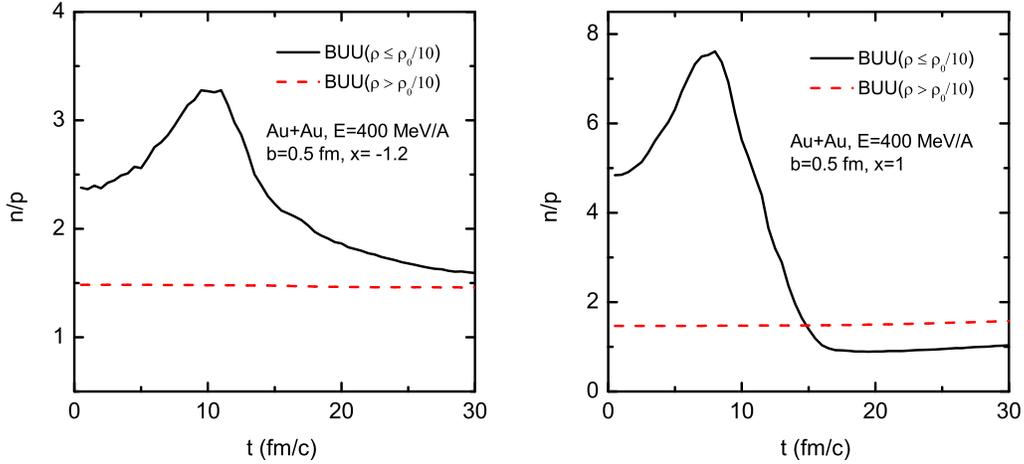}
\caption{Evolution of neutron to proton ratio of free and bound
nucleons in the central $^{197}Au+^{197}Au$ reaction at a beam
energy of 400 MeV/A with positive (left) and negative (right)
symmetry potentials at supra-saturation densities. Simulated with
the IBUU04 transport model ($t_{max}$= 40 fm/c).}
\label{evolution}
\end{figure*}
\begin{figure*}[htb]
\centering\emph{}
\includegraphics[width=0.99\textwidth]{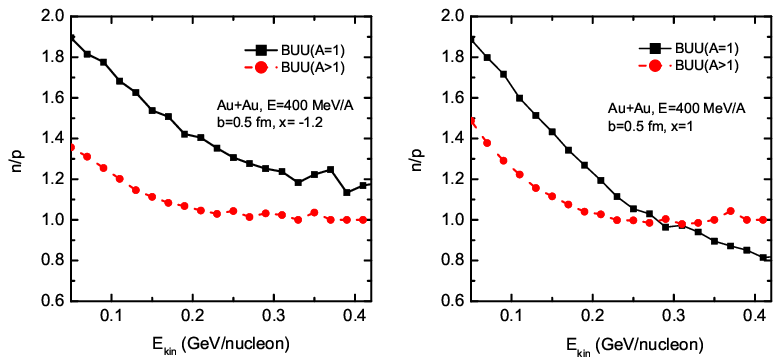}
\caption{Neutron to proton ratio $n/p$ of free and bound nucleons
as a function of nucleon kinetic energy in the central
$^{197}Au+^{197}Au$ reaction at a beam energy of 400 MeV/A with
positive (left) and negative (right) symmetry potentials at
supra-saturation densities. Simulated with the IBUU04 transport
model ($t_{max}$= 40 fm/c).} \label{buukinetic}
\end{figure*}
\begin{figure*}[htb]
\centering\emph{}
\includegraphics[width=0.99\textwidth]{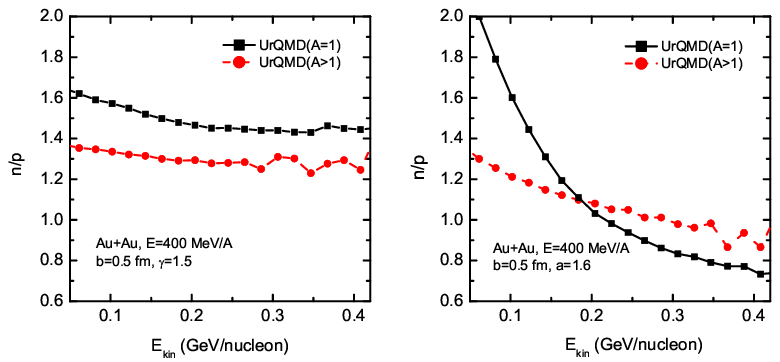}
\caption{Neutron to proton ratio $n/p$ of free and bound nucleons
as a function of nucleon kinetic energy in the central
$^{197}Au+^{197}Au$ reaction at a beam energy of 400 MeV/A with
positive (left) and negative (right) symmetry potentials at
supra-saturation densities. Simulated with the UrQMD transport
model ($t_{max}$= 100 fm/c).} \label{qmdkinetic}
\end{figure*}

The isospin-fractionation (IsoF) during nuclear liquid-gas phase
transition in dilute asymmetric nuclear matter has been studied
extensively \cite{HMuller:95,LiBA:00,liba2007,yong2010}. It is a
common phenomenon that gas phase is more neutron-rich than liquid
phase in dilute asymmetric nuclear matter
\cite{VBaran:2005,Ph:04,LiBA:00}. Here we theoretically confirmed
an abnormal phenomenon that the neutron to proton ratio of gas
phase becomes smaller than that of liquid phase for energetic
nucleons under the action of negative symmetry potential at
supradensities.

With positive and negative symmetry potentials at supradensities
in the used IBUU04 transport model, we first use a simple local
density method to study gas-liquid phase transition. Nucleons with
local density less or equal than $\rho_0$/10 are considered as
nuclear gas, with local density larger than $\rho_0$/10 are
considered as nuclear liquid. Shown in Fig.~\ref{evolution} is
evolution of n/p ratios of nucleons with local density smaller
than $\rho_0$/10 and larger than $\rho_0$/10 from the central
$^{197}Au+^{197}Au$ reaction at a beam energy of 400 MeV/A under
the action of positive and negative symmetry potentials at
supra-saturation densities. We can see that the n/p ratio of gas
phase $(\rho\leq\rho_0/10)$ is larger than that of liquid phase
$(\rho>\rho_0/10)$ for the positive symmetry potential at
supradensities. However, for the negative symmetry potential at
supradensities, an opposite conclusion is obtained. The reason is
that the negative symmetry potential trends to attractive for
neutrons and repulsive for protons during IsoF. Whereas for the
positive symmetry potential, neutrons tend to being repelled by
the symmetry potential and protons tend to being attracted during
IsoF. Thus we get normal and abnormal gas-liquid phase transition
with positive and negative symmetry potentials as shown in
Fig.~\ref{evolution}.

In real nuclear experiments, one in fact gets free or bound
nucleons at final stage. We thus make more realistic predictions
on gas-liquid phase transition, i.e., consider $A= 1$ free
nucleons as nuclear gas whereas $A> 1$ bound nucleons fragments as
nuclear liquid. For the IBUU04 model, we give free and bound
nucleons fragments analysis as that in Ref. \cite{yongcluster}.
Figure~\ref{buukinetic} shows nucleon kinetic energy dependence of
the n/p ratios of free (gas) and bound (liquid) nucleons in the
central $^{197}Au+^{197}Au$ reaction at a beam energy of 400 MeV/A
with positive $(x= -1.2)$ and negative $(x= 1)$ symmetry
potentials at supra-saturation densities simulated by the IBUU04
transport model. From the left panel, we can see that the value of
$n/p$ of nuclear gas phase $(A= 1)$ is larger than that of liquid
phase $(A> 1)$ in the whole kinetic energy distribution region
with the positive symmetry potential $(x= -1.2)$. However, it is
interesting to find that the value of $n/p$ of gas phase $(A= 1)$
is smaller than that of liquid phase $(A> 1)$ at higher kinetic
energies with the negative symmetry potential $(x= 1)$ at
supradensities. This phenomenon in fact can be analyzed and
checked by FOPI-LAND Collaboration \cite{Russotto2011,cozma2013}.

To further confirm normal and abnormal IsoF at higher energies as
previously shown in figure~\ref{buukinetic}, we used the UrQMD
(QMD-like model has advantage over many-body correlation and thus
frequently used to predict cluster production in heavy-ion
collisions \cite{LiQF09}) transport model do the same thing,
which is shown in Fig.~\ref{qmdkinetic}. Again the same phenomenon
occurs, i.e., at higher kinetic energy region, the value of $n/p$
ratio of gas phase $(A=1)$ is becoming smaller than that of liquid
phase $(A> 1)$ with the negative symmetry potential at
supradensities. One discrepancy is that the kinetic energy of
transition point with the UrQMD model is lower than that with the
IBUU04 model. In fact, the kinetic energy of transition point with
the UrQMD model is somewhat $a$ parameter (shown in Eq.~(4))
dependent, but not sensitive to $a$ parameter. To check the
reliability of UrQMD model's predictions, we made the same
simulation but switching off symmetry potential. As expected,
without symmetry potential the values of $n/p$ ratio of gas phase
$(A=1)$ is almost the same as that of liquid phase $(A> 1)$.

Comparing figure~\ref{buukinetic} with figure~\ref{qmdkinetic},
with negative symmetry potential at supradensities, why at low
energy region both IBUU04 and UrQMD give normal IsoF while at
energetic region both show abnormal IsoF? This is because emitted
nucleons with lower energies mainly come from low-density region
of compressed nuclear matter whereas nucleons with higher energies
mainly come from high-density region of compressed nuclear matter,
which are mainly affected by low-density and high-density
behaviors of nuclear symmetry potential, respectively. As shown in
the right panel of figure~\ref{potential}, at low density region
the value of symmetry potential is in fact positive whereas at
high density region it becomes negative. It is the high-density
behavior of nuclear symmetry potential who affects energetic
nucleon emission during IsoF. Therefore energetic nucleon emission
reflects high-density behavior of nuclear symmetry energy. The big
advantage of using this probe is that one just needs to
qualitatively confirm normal or abnormal isospin-fractionation of
energetic nucleons experimentally. Compared with other widely
studied probes in the literature, this confirmation in fact does
not depend on specific numerical values of simulation.


In summary, we provide a qualitative observable, i.e.,
Iso-fractionation for energies nucleons, to give qualitative
determination on nuclear symmetry energy at supra-saturation
densities. With positive/negative symmetry potential (i.e.,
stiff/soft symmetry energy) at supradensities, transport models
give normal/abnormal iso-fractionation for energetic nucleons.
Future accurate experimental measurements (as done by FOPI-LAND
Collaboration) can qualitatively pin down high-density behavior of
nuclear symmetry energy by normal or abnormal Iso-fractionation of
energetic nucleons.


This work is supported by the National Natural Science Foundation
of China (Grant Nos. 11375239, 11435014, 11375062, 11175219,
11175074), the 973 Program of China (No. 2007CB815004), the
Knowledge Innovation Project(KJCX2-EW-N01) of Chinese Academy of
Sciences, the project sponsored by SRF for ROCS, SEM and the
National Key Basic Research Program of China (No. 2013CB834400).

\end{document}